# Thermo-magnetic history effects in the vortex state of YNi$_2$B$_2$C superconductor

**Pradip Das[1], C.V. Tomy[1], H. Takeya[2], S. Ramakrishnan[3] and A.K. Grover[3]**

[1]Department of Physics, Indian Institute of Technology Bombay, Mumbai 400076, India

[2]National Institute of Material Science, Ibaraki 305-0047, Japan

[3]DCMP&MS, Tata Institute of Fundamental Research, Mumbai 400 005, India

E-mail: grover@tifr.res.in

**Abstract.** The nature of five-quadrant magnetic isotherms for $H \parallel a$ is different from that for $H \parallel c$ in a single crystal of YNi$_2$B$_2$C, pointing towards an anisotropic behaviour of the flux line lattice (FLL). For $H \parallel a$, a well defined peak effect (PE) and second magnetization peak (SMP) can be observed and the loop is open prior to the PE. However, for $H \parallel c$, the loop is closed and one can observe only the PE. We have investigated the history dependence of magnetization hysteresis data for $H \parallel a$ by recording minor hysteresis loops. The observed history dependence in $J_c(H)$ across different anomalous regions are rationalized on the basis of superheating/supercooling of the vortex matter across the first-order-like phase transition and possible additional effects due to annealing of the disordered vortex bundles to the underlying equilibrium state.

## 1. Introduction

The order-disorder (O-D) transition in weakly pinned vortex matter and the supercooling/superheating effects associated with it have attracted a large number of studies [1-8] related to the thermomagnetic history effects. These studies reveal anomalous variations in critical current density ($J_c$) that occur in different parts of the ($H,T$) phase space in a variety of superconductors. The most studied anomalous variation occurring close to the $T_c(H)$ line is referred to as the peak effect (PE) phenomenon and is considered to be the consequence of a collapse of elasticity of the flux line lattice (FLL), thereby imbibing some of the attributes (e.g. supercooling/superheating) of the first order FLL melting transition in high $T_c$ cuprates [2]. The second magnetization peak (SMP) anomaly [8] occurring deep into the mixed state and the plateau effect [9] occurring just above the $H_{c1}(T)$ line are the other two often studied anomalous variation in $J_c$, whose connections with the first-order-like phase transition are yet to be firmly established.

In the above context, it is useful to recall a study by Ravikumar et al [8] of the metastability effects in a conventional superconductor V$_3$Si ($T_c$ ~ 17 K) and a high $T_c$ cuprate superconductor

NdBa$_2$Cu$_3$O$_{7-\delta}$ ($T_c$ ~ 95 K), where they showed the presence of an O-D transition at low fields in addition to the PE/SMP anomaly in the low $T_c$/high $T_c$ samples, respectively at high fields. The nature of supercooling/superheating across the low field O-D transition were noted to be somewhat different (i.e., phase-reversed) than those across the high field O-D transition in the sense that the $J_c(H)$ values, while ramping the field down, were smaller than the corresponding values, while ramping the field up after zero field cooling (ZFC). It is pertinent to state that no anomalous variation in $J_c(H)$ was reported to occur across the low field O-D line in both V$_3$Si and NdBa$_2$Cu$_3$O$_{7-\delta}$. High quality single crystals of borocarbide superconductors have emerged as ideal candidates for exploring a variety of novel issues [11,12] in the vortex phase diagram studies. While studying angular dependence of the PE phenomena in a parallelepiped shaped single crystal of YNi$_2$B$_2$C ($T_c$ ~ 15.5 K), we noted additional anomalies in some crystal orientations w.r.t. the applied magnetic field. This motivated us to measure in detail the thermomagnetic history effects of $J_c(H)$ across the PE and the SMP, especially for $H \parallel a$. In particular, we find that the sense of supercooling/superheating across the PE and the SMP region is phase-reversed, like the high field/low field O-D transition in V$_3$Si. Such observations can be rationalized by the premise that the disordered vortex bundles injected into the sample at low fields, sometimes have insufficient opportunity to anneal to the underlying ordered state and hence the residual disorder remained superheated at low fields during the ZFC ramp up of the field [10].

## 2. Experimental

The parallelepiped shaped crystal (3.5x0.70x0.67 mm$^3$) of YNi$_2$B$_2$C [13] was the same crystal that had been employed for point contact spectroscopy studies earlier [14]. The longest dimension of the crystal is parallel to the *a*-axis [100] of the tetragonal structure and the cross-section 0.70x0.67 mm$^2$ comprises [010] and [001] orientations (see figure 1). The magnetization hysteresis studies were performed using a vibrating sample magnetometer (PPMS, Quantum Design USA). Minor hysteresis loops were typically recorded in three modes: (i) after cooling the sample in a given field to 5 K, the field was either ramped down (FC-rev) or up (FC-for), (ii) after cooling the sample in zero field to 5 K, increase the field up to a chosen field and then ramp it down (ZFC-rev) and (iii) after ramping the field down to a pre-selected value at 5 K from above $H_{c2}$, ramp it up again (RL-for).

## 2. Results and discussion

Two premises of the present study are that (i) the magnetization hysteresis under different circumstances can be related to $J_c(H)$ for different thermomagnetic history of applied fields and (ii) $J_c(H)$ relates inversely to a correlation volume $V_c$ ($J_c \propto 1/\sqrt{V_c}$) of the vortex lattice within which the flux lines are collectively pinned [5,8,9]. Figure 1 shows portion of the representative *M-H* loops at 5 K for $H \parallel c$ and $H \parallel a$ (long). The PE is evident in both the directions, however, the SMP can be resolved only for $H \parallel a$. It is interesting to note that the loop is almost closed for $H \parallel c$ prior to the onset field ($H_p^{on}$) of PE, implying a

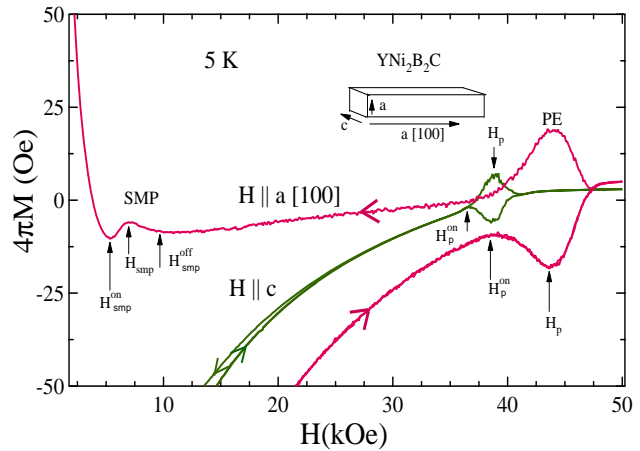

**Figure 1.** Portion of the *M-H* loops at 5 K for $H \parallel a$ and $H \parallel c$ in YNi$_2$B$_2$C.

large change in $V_c$ from $H_p^{on}$ to $H_p$. As asserted by Ravikumar et al [8], if the residual disordered meta-stable phase controls the nucleation of order-disorder transition, the data in figure 1 imply that the co-existing disordered phase is significantly higher prior to the PE for $H \parallel a$ on the forward leg, as compared to that for $H \parallel c$. A natural curiosity is to see whether this disordered phase for $H \parallel a$ is higher on the forward leg or in the reverse leg (due to the appearance of SMP). An ideal way to verify this conjecture is by tracing the minor hysteresis loops. Figure 2 shows traces of minor hysteresis curves at

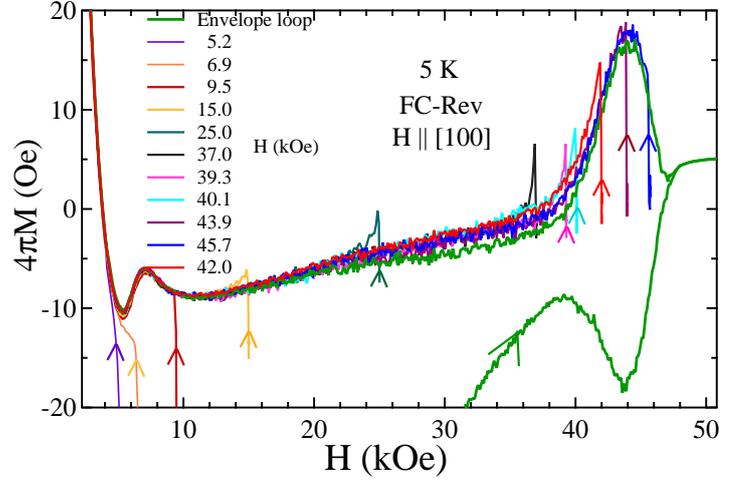

**Figure 2**. Minor hysteresis loops originating from respective field-cooled magnetization values in YNi$_2$B$_2$C for $H \parallel a$ at 5 K.

5 K for $H \parallel a$ in the FC-rev mode. The observation that the $M_{FC\text{-}rev}$ curves overshoot the reverse envelope curve for $H_{smp} < H < H_p$ implies that $J_c^{FC}(H) > J_c^{rev}(H)$, which in turn, corroborates the supercooling of disordered phase in this field region. For $H < H_{smp}^{on}$, $M_{FC\text{-}rev}$ curves undershoot the envelope curve on reverse leg, which could imply that $J_c^{FC}(H) < J_c^{rev}(H)$ in this region.

Figures 3(a) and 3(b) show the traces of minor hysteresis curves originating from field values on the forward leg/ reverse leg and ramped towards lower/ higher field values, respectively, across the PE region. Both sets of minor curves marginally undershoot/ overshoot the respective envelope curves on the reverse/forward legs for $H_p^{on} < H < H_p$. These data imply that the superheating/supercooling effects above/below $H_p^{on}$ are feeble for the vortex states reached along the envelope loop on the forward/reverse legs. Much larger differences were anticipated on the basis of similar data recorded in $V_3Si$ [8], $NdBa_2Cu_3O_{7-\delta}$ [8] and $2H$-$NbSe_2$ [5].

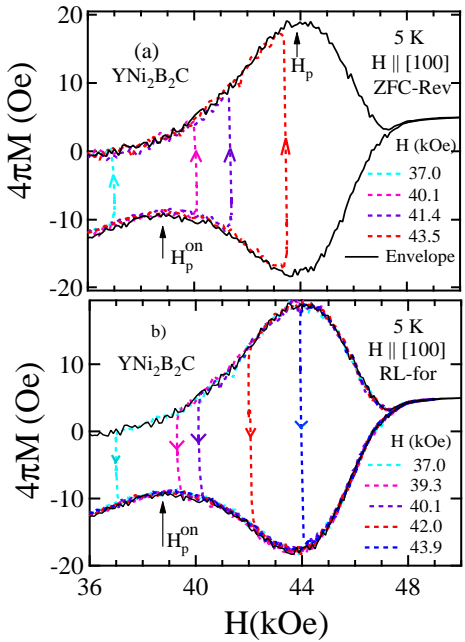

Figure 3. Minor magnetization curves originating from the chosen field values across the onset field of the PE lying on the forward (a) and reverse (b) legs of the envelope hysteresis loop in YNi$_2$B$_2$C for $H \parallel a$ at 5 K.

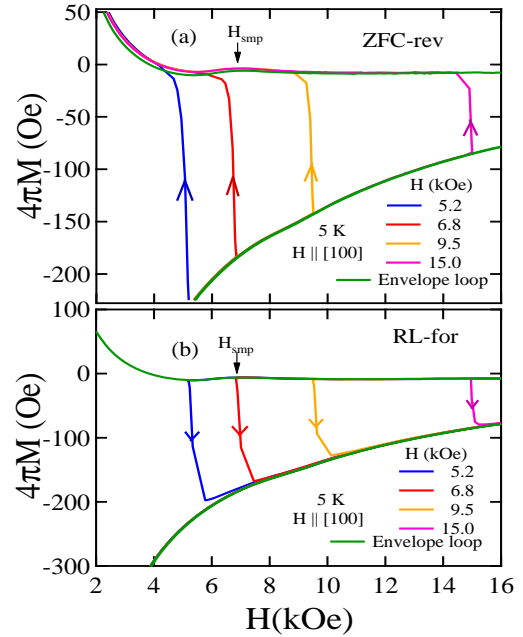

Figure 4. Minor hysteresis curves originating from pre-selected field values across the field region of second magnetization peak anomaly and lying on the (a) forward and (b) reverse legs of the envelope loop in YNi$_2$B$_2$C for $H \parallel a$ at 5 K.

In figuress 4(a) and 4 (b), we now show the minor hysteresis curves initiated from the chosen field values on the forward/ reverse envelope loops across the SMP region. It is interesting to note that the minor curves in figure 4(a)/figure 4(b) marginally overshoot/undershoot the reverse/forward leg of the envelope loop, a behaviour identical to that reported in the crystal $V_3Si$ and $NdBa_2Cu_3O_{7-\delta}$ [8] (even though these crystals did not show the SMP anomaly). However, we can compare the present observation with that of $Ca_3Rh_4Sn_{13}$ ($T_c$ ~ 8 K) in which the SMP anomaly could be indentified distinctly from the PE for $T < 4.5$ K [15].

To conclude, we may state that the notion articulated by Ravikumar et al [8] that $J_c(H)$ values at low fields on the forward leg are larger than those on the reverse leg has a wider applicability. It appears related to the underlying order-disorder transition boundary at low fields, whose fingerprint as an anomalous variation in $J_c(H)$, *a la* second magnetization peak anomaly may, however, be not clearly discernible in the forward leg of the magnetization hysteresis loop, due to an interplay between the rapid decrease of $J_c(H)$ with $H$ at low fields and an annealing process [10] involving settling down of disordered vortex bundles into the underlying stationary state. The thermomagnetic history effects across the PE and the SMP in $YNi_2B_2C$ thereby corroborate the first order character of the underlying order-disorder transition occurring at high and low fields.